# Ground-based detection of an extended helium atmosphere in the Saturn-mass exoplanet WASP-69b

**Authors:** Lisa Nortmann[1,2]*, Enric Pallé[1,2], Michael Salz[3], Jorge Sanz-Forcada[4] , Evangelos Nagel[3] , F. Javier Alonso-Floriano[5], Stefan Czesla[3], Fei Yan[6], Guo Chen[1,2,7], Ignas A. G. Snellen[5], Mathias Zechmeister[8], Jürgen H. M. M. Schmitt[3], Manuel López-Puertas[9], Núria Casasayas-Barris[1,2], Florian F. Bauer[8,9], Pedro J. Amado[9], José A. Caballero[4], Stefan Dreizler[8], Thomas Henning[6], Manuel Lampón[9], David Montes[10], Karan Molaverdikhani[6], Andreas Quirrenbach[11], Ansgar Reiners[8], Ignasi Ribas[12,13], Alejandro Sánchez-López[9], P. Christian Schneider[3], María R. Zapatero Osorio[14]

**Affiliations:**
[1] Instituto de Astrofísica de Canarias, Vía Láctea s/n, 38205 La Laguna, Tenerife, Spain.
[2] Departamento de Astrofísica, Universidad de La Laguna, 38206 La Laguna, Tenerife, Spain.
[3] Hamburger Sternwarte, Universität Hamburg, Gojenbergsweg 112, 21029 Hamburg, Germany
[4] Centro de Astrobiología (Consejo Superior de Investigaciones Científicas - Instituto Nacional de Técnica Aeroespacial, CSIC-INTA), European Space Astronomy Centre campus, Camino bajo del castillo s/n, 28692 Villanueva de la Cañada, Madrid, Spain
[5] Leiden Observatory, Leiden University, Postbus 9513, 2300 RA, Leiden, The Netherlands
[6] Max-Planck-Institut für Astronomie, Königstuhl 17, 69117 Heidelberg, Germany
[7] Key Laboratory of Planetary Sciences, Purple Mountain Observatory, Chinese Academy of Sciences, Nanjing 210008, China
[8] Institut für Astrophysik, Georg-August-Universität, 37077 Göttingen, Germany
[9] Instituto de Astrofísica de Andalucía (Consejo Superior de Investigaciones Científicas, CSIC), Glorieta de la Astronomía s/n, 18008 Granada, Spain
[10] Departamento de Astrofísica y Ciencias de la Atmósfera, Facultad de Ciencias Físicas, Universidad Complutense de Madrid, 28040 Madrid, Spain
[11] Landessternwarte, Zentrum für Astronomie der Universität Heidelberg, Königstuhl 12, 69117 Heidelberg, Germany
[12] Institut de Ciències de l'Espai (Consejo Superior de Investigaciones Científicas, CSIC), Campus Universitat Autònoma de Barcelona, c/ de Can Magrans s/n, 08193 Bellaterra, Barcelona, Spain
[13] Institut d'Estudis Espacials de Catalunya, 08034 Barcelona, Spain
[14] Centro de Astrobiología (Consejo Superior de Investigaciones Científicas - Instituto Nacional de Técnica Aeroespacial, CSIC-INTA), Crta. de Ajalvir km 4, E-28850 Torrejón de Ardoz, Madrid, Spain

*Correspondence to: nortmann.astro [at] gmail [dot] com

## Abstract

**Hot gas giant exoplanets can lose part of their atmosphere due to strong stellar irradiation, affecting their physical and chemical evolution. Studies of atmospheric escape from**

**exoplanets have mostly relied on space-based observations of the hydrogen Lyman-α line in the far ultraviolet which is strongly affected by interstellar absorption. Using ground-based high-resolution spectroscopy we detect excess absorption in the helium triplet at 1083 nm during the transit of the Saturn-mass exoplanet WASP-69b, at a signal-to-noise ratio of 18. We measure line blue shifts of several km s$^{-1}$ and post transit absorption, which we interpret as the escape of part of the atmosphere trailing behind the planet in comet-like form.**

**One Sentence Summary:**

**Studying exoplanets atmospheric escape processes tied to stellar extreme ultraviolet irradiation.**

**In recent years, high-resolution spectroscopy has become a frequently-used tool for investigating exoplanet atmospheres (*1-4*).** Numerous stable high-resolution spectrographs have been deployed on telescopes specifically for exoplanetary science (*5-8*). One of them is the CARMENES (Calar Alto high-Resolution search for M dwarfs with Exoearths with Near-infrared and optical Échelle Spectrographs) spectrograph (*8*) at the 3.5m-telescope of the Calar Alto Observatory. The spectrograph simultaneously covers the visible wavelength range from 0.52 to 0.96 μm and the near-infrared range from 0.96 to 1.71 μm. The near-infrared coverage gives access to exoplanet atmospheric features which cannot be observed in the visible, including the triplet of metastable He I lines around 1083 nm. This feature has been proposed as tracer for atmospheric evaporation (*9*), a process whereby intense X-ray (~0.5 - 10.0 nm) and extreme ultraviolet (EUV, 10.0 - 92.0 nm) irradiation from a host star causes atmospheres of hot gas exoplanets to expand, resulting in a bulk mass flow away from the planet. The continuous mass loss most strongly affects small sub-Neptune sized planets and may be capable of removing their entire volatile atmosphere (*10*). Helium absorption at 1083 nm is sensitive to the low-density gas in an evaporating atmosphere (*9*, *11, 12*) and its observation is not affected by absorption in the foreground interstellar medium, which hampers studies of the neutral hydrogen line Lyman-α (Lyα hereafter) (*9*). He I absorption has been detected in a transmission spectrum of the exoplanet WASP-107b using data from the Hubble Space Telescope (*13*). However, the low resolution prevented a detailed study of the line triplet, including its shape, depth and temporal behavior.

**The Saturn-mass exoplanet WASP-69b orbits an active star with a period of 3.868 days (*14*).** It is an excellent target for atmospheric studies, due to its large atmospheric scale height and high planet-to-star radius ratio, facilitating the detection of a 5.8 ± 0.3% excess absorption in the Na D line (*15*). We used the CARMENES spectrograph to observe two transits of WASP-69b on 2017 August 22 and 2017 September 22 (night 1 and night 2 respectively, see Table S1 for the observing log). The observations spanned approximately four hours for each epoch, which covers the full transit and provides a before- and after-transit baseline. In total 66 spectra were taken, 31 of them out-of-transit.

The wavelength region surrounding the He I feature is affected by emission and water vapor absorption lines originating from within Earth's atmosphere (Fig. S1). Although these lines are spectrally separated from the He I triplet, we correct the effect of water absorption using the European Organisation for Astronomical Research in the Southern Hemisphere (ESO) tool `Molecfit` (*16*), and the sky emission lines using an empirical model derived from the data (*17*). After this correction, we perform continuum normalization and bring the spectra to the stellar velocity rest frame. We then compute a master out-of-transit spectrum ($F_{out}$) which is used to normalize all spectra, following standard analysis methods (*1, 15*). The resulting residual spectra contain the exoplanet absorption signal (Fig. S2). We shift them into the planetary rest frame and compute the transmission spectrum by co-adding all residual in-transit spectra ($F_{in}/F_{out}$) obtained between second and third contact (see Fig. 1), i.e. when the planet disk was fully in front of the stellar disk.

**The combined transmission spectrum for the two nights is shown in Figure 2. An excess absorption in the He I line at the level of 3.59% ± 0.19% is detected.** The given uncertainty corresponds to one standard deviation (1σ) of the continuum flux. The signal is detected separately in each visit at 3.96% ± 0.25% (1σ) and 3.00% ± 0.31% (1σ) for nights 1 and 2, respectively (Fig. S3). We model the transmission spectrum with three Gaussian functions with fixed amplitude ratios and relative wavelengths according to theoretical values for the He I triplet (*18, 19*). We fit a common line width, Doppler-shift, and intensity of the lines (*17*) and determine parameter uncertainties by Markov Chain Monte Carlo sampling (Fig. S4). The best

fitting model indicates a net blue shift of $-3.58 \pm 0.23$ km s$^{-1}$ (where the uncertainty corresponds to the standard deviation of the posterior probability distribution).

To examine the behavior of the helium absorption over time, we construct a light curve by summing the flux within a 0.04 nm passband centered on the blue shifted core of the He I feature for each residual spectrum in the planet rest frame (*15*). The resulting light curves for each of the two nights are shown in Figure 3. The helium absorption begins shortly after the planet ingress, with no observable pre-transit absorption, and lasts for 22 ± 3 min after the transit ends (see Fig. S5). This light curve behavior does not depend on the width of the chosen passband. By fitting the Rossiter-McLaughlin effect (RME), a deformation of the stellar lines caused by the planet occulting different parts of the rotating stellar surface during transit, in our visible channel radial velocity data (*17*, Fig. S6) we obtain mid-transit times consistent with the known planet orbit. The signal of the RME corresponds with the predicted broadband transit duration of 2.23 h (*14*), so we can be confident that the observed post-transit helium absorption is real. We use the RME curve to estimate the potential contamination of the transmission spectrum by the corresponding deformation of the stellar lines during transit; we find the impact is negligible (*17* and Fig. S7). The He I $D_3$ line at 587.6 nm and the Ca II infrared triplet (IRT) at 849.8, 854.2 and 866.2 nm, both indicators of stellar activity, show no sign of active regions (*17* and Fig. S8). The time delay of the helium ingress and egress indicates that the distribution of helium around the planet is asymmetrical and that a cloud of gas is trailing the planet along its orbit (Fig. 1). We calculate the length of this tail as approximately 170,000 km, i.e. 2.2 times the planet radius (longer if tilted with respect to the planet's orbit). Acceleration of the tail material away from the planet could be the cause of the blue shifted absorption. This hypothesis is supported by the larger measured net blue shift of $-10.69 \pm 1.00$ km s$^{-1}$ when only the helium tail is occulting the stellar disk (Fig. S9). The tail length and velocities suggest that helium is escaping the planet (*17*).

**We also analyzed CARMENES transit observations of the hot Jupiter-mass exoplanets HD 189733b and HD 209458b, the extremely hot planet KELT-9b, and the warm Neptune-sized exoplanet GJ 436b (Fig. S10).** GJ 436b and HD 209458b both show evaporation of hydrogen in the Lyα line (*20, 21*) and KELT-9b is surrounded by a large cloud of evaporating hydrogen absorbing in the Balmer Hα line at 656.28 nm (*22*). GJ 436b and

HD 209458b are predicted to have large absorption depths in the He I line (~8% and ~2%, respectively) (*9*), although a previous study of HD 209458b did not detect any absorption (*23*). We do not detect He I absorption in most of these planets, with 90% confidence upper limits of 0.41% for GJ 436b, 0.84% for HD 209458b, i.e. in disagreement with the predicted levels (*9*), and 0.33% for KELT-9b (Fig. S10). However, we do detect helium absorption in HD 189733b at the level of 1.04% ± 0.09 (*24*). A companion paper reports a similar detection of helium absorption for the warm Neptune-sized planet HAT-P-11b (*25*). For our detections we calculate the equivalent height of the He I atmosphere $\delta_{Rp}$, i.e. the height of an opaque atmospheric layer that would produce the observed absorption signal (see Table S2). For both WASP-69b and HD 189733b we find $\delta_{Rp}$ to be ~80 times larger than the atmospheric scale height $H_{\mathrm{eq}}$ calculated for the respective planet's deep atmosphere i.e., in hydrostatic equilibrium (*17*). For the other three planets our upper limits correspond to no detections of features above ~40 $H_{\mathrm{eq}}$.

**Why do similar hot gas exoplanets show such a range of helium absorption values?** The expansion of the escaping planetary atmosphere depends on parameters like the extreme ultraviolet irradiation and the planetary density (26), but the population of the helium triplet state depends on the irradiation at wavelengths < 50.4 nm (*9*). While GJ 436b and HD 209458b orbit very quiet stars (*27, 28*), the hosts of the planets in which helium is detected, i.e. WASP-69, HD 189733, HAT-P-11 and WASP-107, are all relatively active stars (*15*, *29*, *30*, *14*). In Figure 4A, we plot the normalized absorption altitude of helium $\delta_{Rp}/H_{\mathrm{eq}}$ against the stellar activity index log($R'_{HK}$) (*31*). Our sample size is limited, but the detections succeeded for the planets with the more active stellar hosts, hinting at a relation between He I detectability and host star activity.

Low-mass stars (F-, G-, K- and M-types) stars have a convective layer that, in combination with stellar rotation, produces phenomena associated with magnetic activity. The exterior layers of low mass stars are (from inside to outside): photosphere, chromosphere, transition region, and corona. In general, activity in the chromosphere is detected in spectral features such as the activity indicator Ca II H and K doublet lines at 393.4 and 396.8 nm, while the transition region and the corona produce emission in X-ray and EUV. The metastable $2^3$S helium triplet state, which is the lower level of the observed absorption lines, is populated via radiative ionization of He I by photons with wavelengths < 50.4 nm followed by recombination (*32*). Thus, a higher X-ray and EUV (< 50.4 nm, hereafter $\mathrm{XUV_{He}}$) irradiation level should enhance the formation of the He I triplet in atmospheres of hot gas planets. We calculated the $XUV_{He}$ flux received by all discussed planets (Table S3) at the separation of their orbit (*17*, Table S4). In Figure 4B, we plot the normalized He I atmospheric altitude $\delta_{Rp}/H_{eq}$ for our measurements as a function of the $XUV_{He}$ flux. The line is detected in the planets receiving the largest combined $XUV_{He}$ irradiation. These results indicate a dependence of the detectability of He I in planetary atmospheres on intense X-ray and EUV emission from the parent star.

**Acknowledgements:**

Parts of the results shown are based on observations obtained with XMM-Newton, an ESA science mission with instruments and contributions directly funded by ESA Member States and NASA. We acknowledge the XMM-Newton Project Scientist for the quick and positive reaction to our request for a director's discretionary time observation of WASP-107. The authors thank the anonymous reviewers for their contribution to this paper. **Funding:** CARMENES is an instrument for the Centro Astronómico Hispano-Alemán de Calar Alto (CAHA, Almería, Spain). CARMENES is funded by the German Max-Planck-Gesellschaft (MPG), the Spanish Consejo Superior de Investigaciones Científicas (CSIC), the European Union through FEDER/ERF FICTS-2011-02 funds, and the members of the CARMENES Consortium (Max-Planck-Institut für Astronomie, Instituto de Astrofísica de Andalucía, Landessternwarte Königstuhl, Institut de Ciències de l'Espai, Insitut für Astrophysik Göttingen, Universidad Complutense de Madrid, Thüringer Landessternwarte Tautenburg, Instituto de Astrofísica de Canarias, Hamburger Sternwarte, Centro de Astrobiología and Centro Astronómico Hispano-Alemán), with additional contributions by the Spanish Ministry of Economy, the German Science Foundation through the Major Research Instrumentation Programme and DFG Research Unit FOR2544 "Blue Planets around Red Stars", the Klaus Tschira Stiftung, the states of Baden-Württemberg and Niedersachsen, and by the Junta de Andalucía. We acknowledge funding from the Spanish Ministry of Economy and Competitiveness (MINECO) and the Fondo Europeo de Desarrollo Regional (FEDER) through grants: ESP2016-80435-C2-1-R, ESP 2016-76076-R, ESP2014-54362-P, ESP 2014-54062-R, AYA2016-79425-C3-2-P, AYA2016-79425-C3-1-P, AYA2016-79425-C3-2-P, AYA2014-54348-C3-1-R and AYA2016-79425-C3-3-P. Further we acknowledge funding through the Deutsche Forschungsgesellschaft (DFG) through: DFG DR281/32-1, RE 1664/14-1, DFG SFB 676, DFG SCHM 1032/57-1 and by the Deutsches Zentrum für Luft- und Raumfahrt (DLR) through grants DLR 50 OR 1710, DLR 50 OR 1307, BMWi50OR1505 as well as the support of the Generalitat de Catalunya/CERCA programme. I.A.G and F.J.A-F. acknowledge funding from the research programme VICI 639.043.107 funded by the Dutch Organisation for Scientific Research (NWO), and funding from the European Research Council (ERC) under the European Union's Horizon 2020 research and innovation programme under grant agreement No 694513. G.C. acknowledges the support by the National Natural Science Foundation of China (Grant No. 11503088) and the Natural Science Foundation of Jiangsu Province (Grant No. BK20151051). **Author contributions:** L.N. performed data acquisition including proposal writing and preparation of observations (DAQ), data analysis (DA) and science interpretation (SI). F.J.A.-F., F.Y. I.A.G.S. preformed DAQ, DA and SI. E.P. and S.C. performed DAQ and SI. M.S., M.L.P., A.S.L., D.M., K.M., N.C.B. performed DA and SI. J.S.F. performed X-ray and EUV flux modeling and calculation, DAQ and SI. E.N. performed telluric absorption line modeling and correction, M.Z. DA, instrument design (ID), official instrument pipeline design (PD), F.F.B. performed DA and PD, A.R. performed DA, ID, PD and SI. I.R. ID, DAQ and SI. A.Q. performed ID, SI. J.A.C. performed ID and DAQ. G.C. performed planet scale height calculations, DAQ and SI. J.H.M.M.S., P.J.A. S.D., T.H., M.L. K.M., C.S. and M.R.Z.O. performed SI. **Competing interests:** The authors have no competing interests to declare. **Data and materials availability:** The results shown in this manuscript are based on data from the CARMENES data archive at Centro de Astrobiología (INTA-CSIC). They can be found in the Calar Alto (CAHA) archive at http://caha.sdc.cab.inta-csic.es/calto/jsp/searchform.jsp using the identifiers given below and specifying the observation dates: 'WASP-69': 22-23.Aug.2017 (CAHA_IDs 261990-262072), 22.Sep.2017 (CAHA_IDs 263371-263467. 'KELT-9b': 06-07.August.2017 (CAHA_IDs 259314-259424). 'HD209458': 16-17.Sep.2016 (CAHA_IDs 249262-249415) and 08.11.2016 (CAHA_IDs 251027 251198). The CARMENES data for GJ 436 is still within proprietary time, therefore, not in the CAHA archive. The data of the echelle order 56, containing the He I line can be found here (http://carmenes.cab.inta-csic.es/gto/jsp/nortmannetal2018.jsp). XMM Newton data can be found in the archive at http://nxsa.esac.esa.int/nxsa-web/#home. The observation IDs used are given in the supplementary material. Scripts written for this manuscript can be found as supplementary files.

**Supplementary Materials:**
Materials and Methods
Figures S1-S10
Tables S1-S4
References (*33*-*74*)

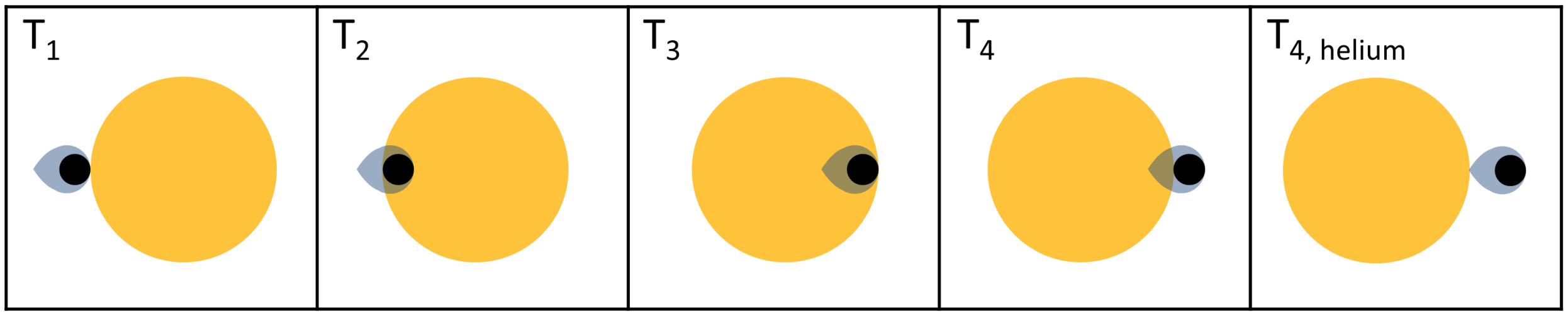

**Fig. 1. Illustration of the exoplanet WASP-69b (black) and its extended helium atmosphere (light blue) at the different contact points.** Shown are the first ($T_1$), second ($T_2$), third ($T_3$) and fourth ($T_4$) contact of the broadband planet transit and the moment when also the tail has passed the stellar disk, $T_{4,helium}$ 22 ± 3 min after $T_4$.

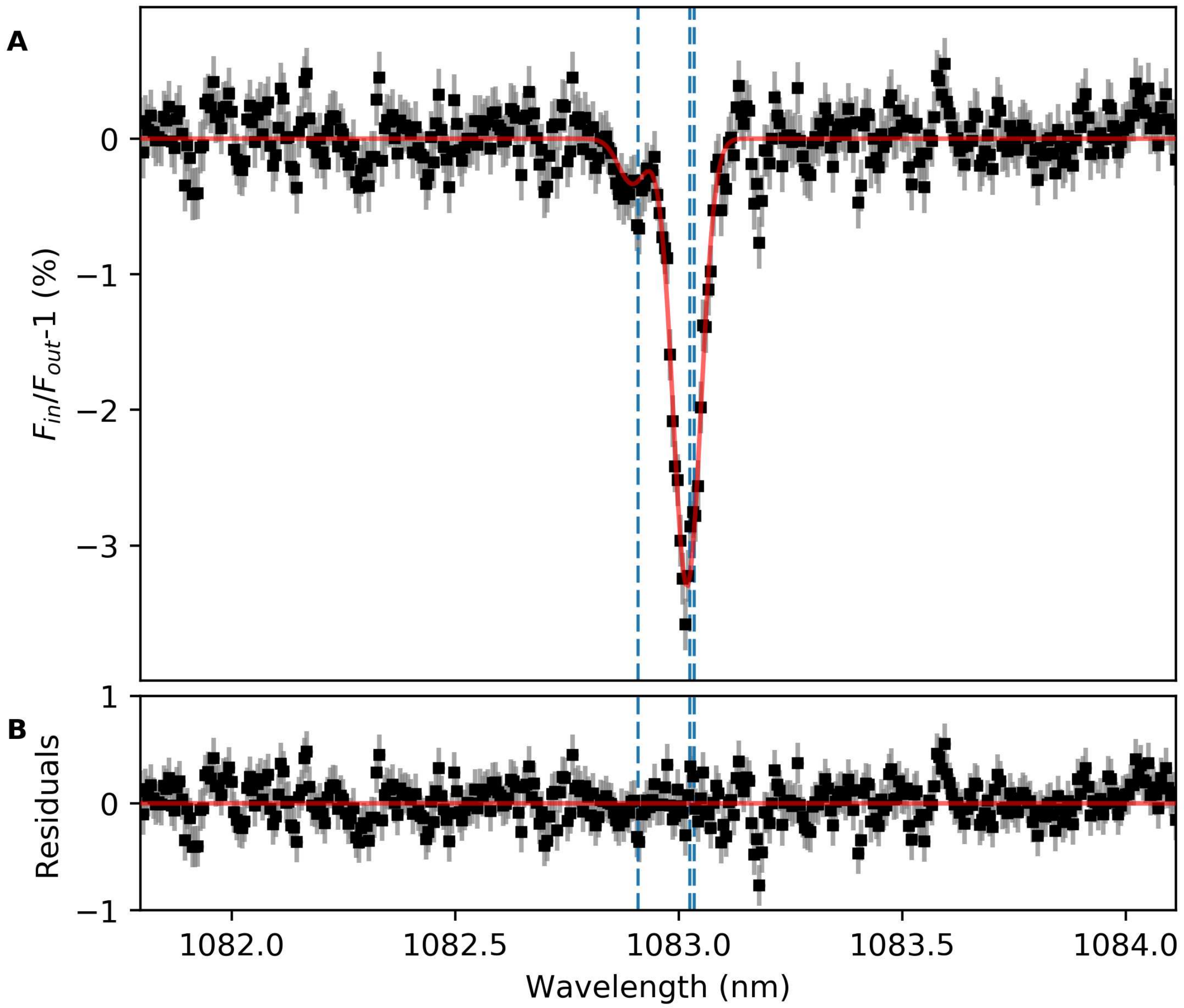

**Fig. 2. Transmission spectrum between second and third contact (see Fig. 1) of WASP-69b showing planetary absorption in the He I triplet at 1083 nm.** A) the excess absorption of helium in the weighted mean averaged transmission spectrum (black points) from two transit observations of WASP-69b (2017 August 22 and 2017 September 22). The best fitting model (red line) shows a net blue shift of $-3.58 \pm 0.23$ km s$^{-1}$. The predicted positions of the helium triplet lines (1082.909 nm, 1083.025 nm and 1083.034 nm) are indicated as vertical dashed blue lines. B) the residuals of the data after subtraction of the model are shown in black, the red line indicates the zero level.

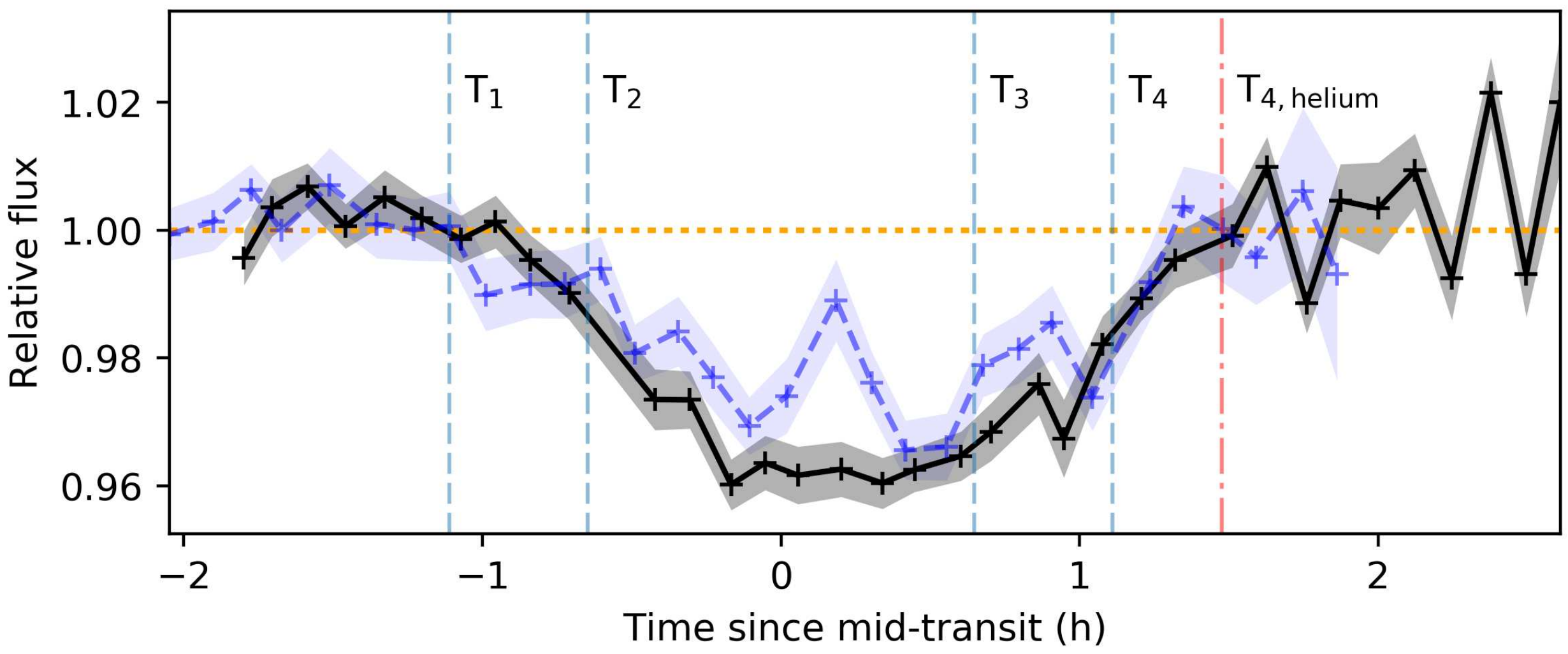

**Fig. 3. Spectrophotometric transit light curves of WASP-69b**. We integrated the spectral flux in a 0.04 nm wide bin around the core of the planetary He I line for every observed spectrum over two transits, normalized by the continuum flux outside of the absorption feature. The first ($T_1$), second ($T_2$), third ($T_3$) and fourth ($T_4$) contact of the planet transit are marked by dashed vertical lines. Two individual transit light curves are shown in black (night 1) and blue (night 2). The drop in flux from the continuum transit has already been removed, leaving the excess absorption due to helium. The continuum behavior is indicated by the horizontal yellow dotted line. 1σ uncertainty intervals are shown as light blue and grey shaded regions. The excess absorption lasts until well after the stellar occultation by the planet has ended ($T_4$), indicating that absorbing material is still in front of the star. We find the excess absorption ends 22 ± 3 min after the planet's egress ($T_{4,\mathrm{helium}}$, vertical red dash-dotted line).

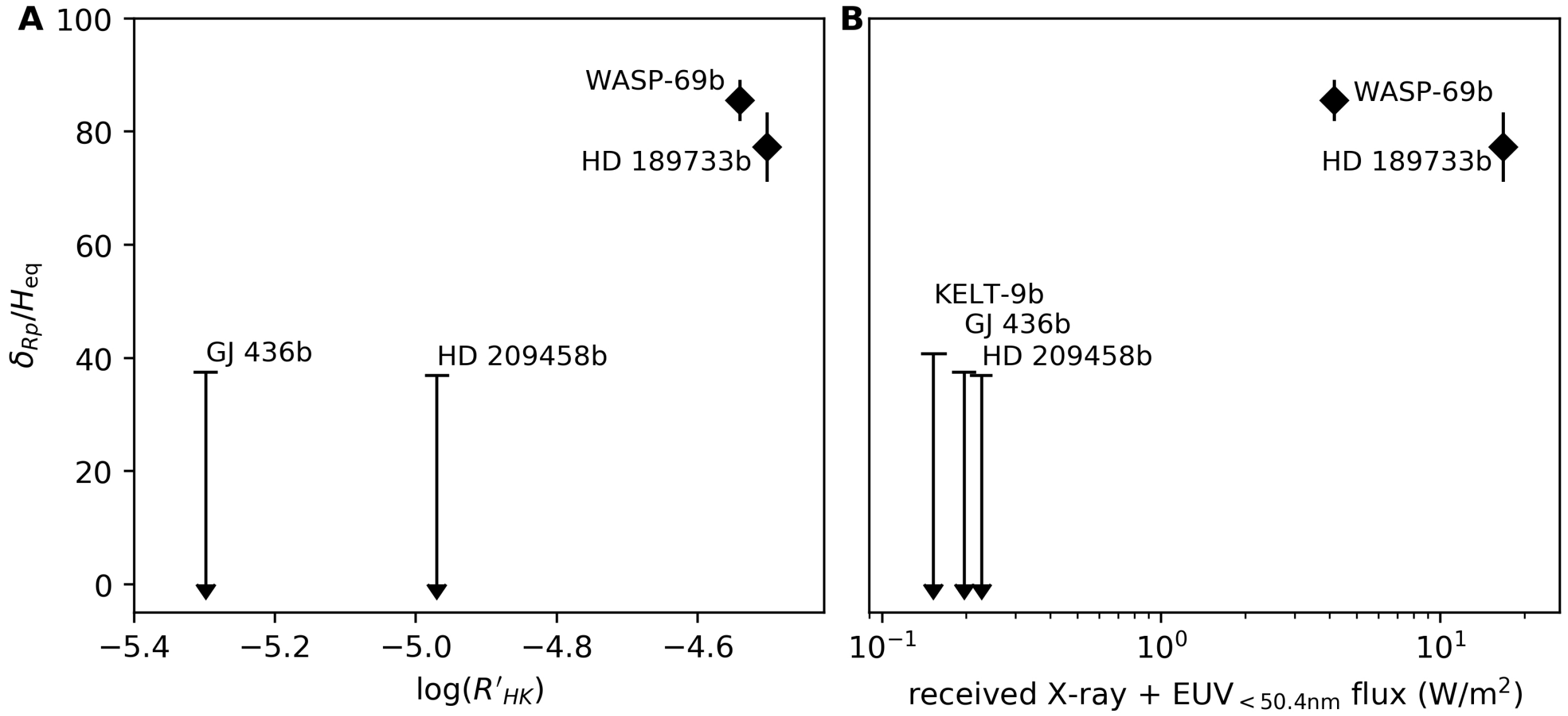

**Fig. 4. Detected signals as a function of host star activity and received $XUV_{He}$ irradiation.** We show the equivalent height of the He I atmosphere $\delta_{Rp}$ normalized by one atmospheric scale height of the respective planet's lower atmosphere $H_{eq}$. For the two detections we plot 1σ error bars, for the non-detections we plot upper limits corresponding to a 90% confidence level. A) $\delta_{Rp}/H_{eq}$ as a function of the host star activity index log($R'_{HK}$), where larger values indicate stronger stellar activity (*31*). The KELT-9 system is not plotted as its log($R'_{HK}$) is not known. In panel B we show $\delta_{Rp}/H_{eq}$ as a function of stellar flux with wavelength < 50.4 nm at the distance of the planet orbit. The two strong detections of an extended helium atmosphere occur for the two planets having more active host stars and higher planetary $XUV_{He}$ irradiation.

# Supplementary Materials for

## Ground-based detection of an extended helium atmosphere in the Saturn-mass exoplanet WASP-69b

**Authors:** Lisa Nortmann[1,2]*, Enric Pallé[1,2], Michael Salz[3], Jorge Sanz-Forcada[4] , Evangelos Nagel[3] , F. Javier Alonso-Floriano[5], Stefan Czesla[3], Fei Yan[6], Guo Chen[1,2,7], Ignas A. G. Snellen[5], Mathias Zechmeister[8], Jürgen H. M. M. Schmitt[3], Manuel López-Puertas[9], Núria Casasayas-Barris[1,2], Florian F. Bauer[8,9], Pedro J. Amado[9], José A. Caballero[4], Stefan Dreizler[8], Thomas Henning[6], Manuel Lampón[9], David Montes[10], Karan Molaverdikhani[6], Andreas Quirrenbach[11], Ansgar Reiners[8], Ignasi Ribas[12,13], Alejandro Sánchez-López[9], P. Christian Schneider[3], María R. Zapatero Osorio[14]

*Correspondence to: nortmann.astro [at] gmail [dot] com

**This PDF file includes:**

Materials and Methods
Figs. S1 to S10
Tables S1 to S4

## Materials and Methods

Observations

We observed WASP-69b with the CARMENES spectrograph during two transit events. Observations were started at low airmass of 1.97 in both nights, 44 min and 56 min before transit ingress in night 1 and night 2, respectively. The target passed its highest point (the meridian) approximately in the middle of the observations and was setting when the observations were stopped in both nights, 92 min (night 1) and 49 min (night 2) after transit egress. As a consequence, spectra obtained towards the end of the night have a lower signal-to-noise ratio compared to the rest of the spectra. A summary of the observations is given in Table S1. Both the visible light (VIS) and near infrared (NIR) channels of the instrument were taking observations quasi-simultaneously. Fiber A was pointed at the star and fiber B was pointed at the sky. The exposure time was chosen as 398 s, so that the planet velocity would change less than 1 km $s^{-1}$ during one exposure, and the signals of narrow lines were not smeared out over multiple pixels. The atmospheric dispersion corrector at the telescope entrance was manually reset during detector readout approximately every 30 min. In total 66 spectra were taken, 35 of these in transit, 16 during ingress and egress and 19 between the second and third contact, i.e. when the whole planetary disk was in front of the host star.

Extraction of spectra, timestamps and radial velocities:

The detector raw frames are processed with the CARMENES pipeline `CARACAL` (*33*) based on flat-relative optimal extraction (*34*) and a wavelength calibration algorithm described in (*35*). Radial velocities (RVs) are measured with the least-squares fitting program `SERVAL` (*36*) and corrected for barycentric motion (*37*).

We find some specific pixels at fixed positions in the extracted spectra to behave differently from the rest of the signal, especially for low signal-to-noise observations. Investigating the raw frames we find these artifacts correspond to hot pixels on the detector. These are not taken into account by the current version of `CARACAL`, so we mask these pixels in the following steps of the analysis. Around 1% of pixels in the extracted spectra are masked.

Correction of telluric lines in the He I region

The He I feature at 1083 nm falls on echelle order 56 in the CARMENES NIR channel. Telluric emission and water absorption lines contaminate the wavelength interval surrounding the feature. These lines need to be corrected before any further steps are taken, as their contribution changes with time and airmass and their position in the spectrum does not move together with the stellar lines, which shift from red to blue shifted over the duration of a transit event.

Water absorption lines: We carry out a correction of telluric water vapor lines in each individual spectrum using the ESO sky software tool `Molecfit` in version 1.2.0 (*16, 38*). `Molecfit` derives synthetic high-resolution atmospheric transmission spectra using a line-by-line radiative transfer model that requires an atmospheric profile as input. The atmospheric profile within `Molecfit` is based on a combination of three different sources: A reference atmosphere profile (*39*), Global Data Assimilation System (GDAS) profiles, and local meteorological data gathered during the observations. For our analysis, we use a reference profile optimized for nightly mid-latitudinal locations at 45°. The GDAS data are provided on a geographical 1° × 1° grid on a time base of 3h and we used the profiles matching the location of the Calar Alto Observatory at the time of the observations. Apart from the atmospheric profile, the line-by-line radiative transfer model requires a transmission molecular absorption database. `Molecfit` is distributed with the database AER version 3.2 (*40*), which is based on the high-resolution transmission molecular absorption database (HITRAN) 2008 (*41*). We perform the model fitting for narrow wavelength ranges that cover unsaturated telluric absorption lines with various strengths. While the atmospheric water vapor content was a free fit parameter, we fix the abundance of other molecules that are present in the CARMENES near infrared spectral range ($O_2$, $O_3$, and $CH_4$) to the value of the reference profile. The fitting intervals are evenly distributed along the entire wavelength range and cover several molecular absorption bands leading to an accurate telluric correction regarding the wavelength calibration (*38*). The continuum level of the model within each fitting range was linearly scaled to match the data. We use the latest generation of synthetic high-resolution PHOENIX spectra (*42*) to exclude stellar features within these ranges. For WASP-69, we used a spectral model with effective stellar temperature $T_{\mathrm{eff}}$ = 4700 K, logarithm of the gravitational acceleration log $g$ = 4.5 dex, and solar metallicity, the parameter combination in the PHOENIX grid closest to literature values (*14*). The model was broadened using a Gaussian kernel to match the resolution of CARMENES. We applied rotational broadening, normalized to the continuum level, and shifted the PHOENIX model to match the wavelength of the observation. Wavelength intervals in the observed spectra that coincide with absorption lines present in the model spectrum were excluded.
Emission lines: The He I feature is bracketed by sky emission lines to left and right. As fiber B of the instrument was pointed at the sky, we have independent observations of the lines for each exposure to correct for this effect. However, we achieve a more precise correction by using an empirical model of the skylines derived from the out-of-transit data. In this empirical model, we set all values outside of the emission line features to zero to prevent introducing noise to the data. The sky emission line windows not set to zero are 1082.894-1082.997 nm, 1083.094-1083.163 nm and 1084.123-1084.207 nm (wavelengths given in air and in telluric rest frame). We then allow this model to scale to the data, independently for each emission feature and subtract the best fit model from the spectra.

Calculation of the transmission spectrum

After we correct telluric absorption and emission, we normalize each spectrum by the continuum. For this, we divide each spectrum by the first spectrum of the night, mask the He I triplet and all regions previously affected by telluric emission and absorption and fit a first order polynomial function to the remaining points. Then we extrapolate this polynomial also to the masked regions and divide each spectrum by it. The masking in this step affects a little less than half of the available spectral range. After the normalization we calculate the projected velocity of the star towards the observer for the mid-exposure time and the corresponding wavelength solution for each spectrum in the stellar rest frame. Then we rebin all spectra to the same wavelength grid using the IDL routine `rebinw` from the PINTofALE package, which ensures flux conservation (*43*). We calculate a master out of transit spectrum for both nights as the mean averaged spectrum of the before and after transit spectra. In this process we exclude the last two spectra of night 1 and the last three spectra of night 2 as they are of lower signal-to-noise. Doing so does not change the overall outcome of the transmission spectrum but it reduces the continuum noise. We then divide all spectra by this out-of-transit master stellar spectrum.
The residual spectra contain the planet transmission spectrum, Doppler-shifted by a different value in each exposure (Fig. S2). This shift is caused by the projected planet velocity with respect to the stellar rest frame changing over the duration of the planet transit (i.e. the planet moves toward the observer until mid-transit and away from the observer afterwards, covering a velocity range of −9.2 to 9.2 km $s^{-1}$). We calculate the planet's projected velocity (*44*) and bring the residual spectra to the same wavelength sampling in the planet rest frame. We then co-add all in-transit spectra (between second and third contact, $T_2$-$T_3$) to obtain the transmission spectrum for both nights. We find an excess absorption signal of He I at a signal-to-noise ratio of 15.8 and 9.7 for night 1 and night 2, respectively. We use the standard deviation of the continuum σ to estimate the noise. We then combine the transmission spectra by calculating the weighted mean of both nights, using $w=1/\sigma^2$ as the respective weights. The two individual and the combined transmission spectra are shown in Fig. S3.

Description of the model for the transmission spectrum

We fitted a model to the weighted mean averaged transmission spectrum using the IDL implementation `mpfit` of the Levenberg–Marquardt least-squares algorithm (*45*). Assuming an optically thin environment, the model consists of three Gaussian functions with fixed amplitude ratios and positions of the He I triplet (*18, 19*). We allow the root mean square (RMS) width (i.e. the standard deviation) of all three Gaussians to vary jointly and, further, treat the amplitude of the deepest line and a velocity causing a Doppler-shift of the lines as free parameters. The uncertainties of the data points used in the Levenberg–Marquardt algorithm were determined as the standard deviation of the transmission spectrum outside of the helium feature. In order to retrieve uncertainty estimates, we determine the shape of the probability distribution of the model parameters with the Markov Chain Monte Carlo (MCMC) method using an affine-invariant ensemble sampler as implemented in the Python package emcee (*46*). The procedure aims to maximize the logarithmic likelihood function described in (*47*). We use non-informative priors for all parameters and, in addition to the model parameters, allow for a factor to scale data uncertainties to fit the true noise of the data. This factor converges to 0.99 ± 0.03. The correlation plots of the parameter posterior distributions are shown in Fig. S4. The posteriors are close to normally distributed without unexpected correlations. The anti-correlation between the parameters for the fitted width and amplitude is expected. As we are using non-normalized Gaussian functions the width and amplitude of the individual modeled lines are not correlated. However, since the deepest point of the triplet feature is made up of two blended lines, increasing the line width (while maintaining the line amplitudes) increases the amount of blending and, thus, the allover feature depth. This increased feature depths then needs to be compensated by lower amplitudes of the individual lines.

The best fitting values adopted from the Levenberg–Marquardt least-squares algorithm and the error bars calculated from the 1σ percentiles of the MCMC posterior parameter distributions are −3.58 ± 0.23 km $s^{-1}$ for the blue shift, 0.0307 ± 0.0008 nm for the common RMS width of the Gaussians and −0.0222 ± 0.0005 for the amplitude of the deepest line of the triplet (rest wavelength 1083.034 nm), with the other two lines scaling proportionally to this value following the theoretical predictions (*18, 19*). We repeat the same analysis for a 'tail' spectrum constructed from the two spectra of each night obtained close in time to the fourth contact ($T_4$), i.e. the moment when the planet has cleared the stellar disk (the used spectra are marked in Fig. S5) and an 'ingress' spectrum constructed from the spectra obtained between $T_1$ and $T_2$ (see Fig. S9 for the spectra and corresponding best fit models). For the 'tail' transmission spectrum we measure a net blue shift of −10.7 ± 1.1 km $s^{-1}$, a common RMS width of the Gaussians of 0.035 ± 0.004 nm and an amplitude of −0.011 ± 0.001 for the deepest line. For the 'ingress', i.e. the $T_1$-$T_2$ transmission spectrum we measure net red shift of 1.4 ± 0.9 km $s^{-1}$, a common RMS width of the Gaussians of $0.015^{+0.007}_{-0.003}$ nm and an amplitude of $-0.007^{+0.001}_{-0.002}$ for the deepest line.

Determination of the He I egress time and tail length

We obtain an estimate for the He I egress time by fitting a theoretical light curve model (*48, 49*) to the excess absorption of both nights shown in Figure 3, simultaneously. We allow the planet-to-star radius ratio, the limb darkening and the mid-transit time to differ from the literature broadband values. We further allow a scaling factor that regulates the depth of the curve independent from the transit duration for both nights. The best fitting mid-transit time of the excess absorption curve converges to 15 ± 1 min after the broadband mid-transit time. Further, the model reproduces the prolonged transit duration. The timing of the fourth contact of the model lags behind from the broadband transit egress by 22 ± 3 min and the timing of the first contact lags behind by 9 ± 3 min. Using the known parameters of the system (*14*) we calculate the planet's orbital velocity $v_{orb} = 2\pi\, a_p / P$, where $a_p$ is the planet's semi-major axis and $P$ is the orbital period. We derive $v_{orb}$= 127.3 ± 1.5 km $s^{-1}$ and, consequently, that the planet travels 170 ± 23 × $10^3$ km, i.e. 2.2 ± 0.3 planet radii in 22 ± 3 min, the duration of the delayed egress, and 70 ± 23 × $10^3$ km i.e. 0.9 planet radii in 9 ± 3 min, the time of ingress delay. Uncertainties are derived by Gaussian error propagation.

Comparison to theoretical studies of atmospheric escape

The velocities of −3.58 ± 0.23 km $s^{-1}$ and −10.69 ± 1.00 km $s^{-1}$ derived for the mid-transit ($T_2$-$T_3$) and tail ($T_4$) spectra, respectively, are about one order of magnitude smaller than peak velocities measured for hydrogen escape in other planets, which are typically 100 to 200 km $s^{-1}$ (*20, 21, 50*). They are also lower than the planetary escape velocity $v_{esc}$ = 29.5 km $s^{-1}$ of WASP-69b. However, the velocity measured at $T_4$ is comparable to wind velocities expected from theoretical studies of atmospheric escape, which are 8 – 12 km $s^{-1}$ (*51*). The tail length suggests that we measure helium at a distance of at least 3.2 $R_p$ from the planet center, which is slightly larger than the extent of the Roche lobe (*52*) in orbital plane, 2.73 $R_p$. This value is calculated as the distance in the orbital direction from the planet center to the equipotential that crosses the $L_1$ Lagrangian point, derived from the effective gravitational potential (*53*, their equation 20). As the measured net blue shifts are likely to correspond to an average velocity of the material, with some parts exhibiting higher and some lower velocities, it is reasonable to assume that some of the helium atoms in the tail can escape the gravitational field of the planet. As the $T_2$-$T_3$ spectrum contains signals from material with different projected velocities towards the observer, this results in a slight broadening of the measured net signal at mid-transit. We, however, do not observe a signal broadening as extreme as 3D models predict for a comet-like tail scenario in the super-Neptune WASP-107b (*13*), nor asymmetry in the line wings.

Fitting of the Rossiter-McLaughlin effect:

The CARMENES spectra are affected by the Rossiter-McLaughlin effect (RME) (*54*). Because we used fiber B to monitor the sky background (instead of alternative monitoring of a Fabry-Perot etalon), we lack simultaneous measurements of the instrument drift. In the VIS channel, however, this drift is small in both nights and in first approximation linear in time, so that the typical pattern of the RME can be directly seen in the RV data. The data was obtained from the spectra using the least-squares fitting program `SERVAL` (*36*), taking into account all the entire wavelength range between 561.2 and 932.5 nm. Shorter and longer wavelengths covered by the VIS channel were excluded due to exhibiting a low signal-to-noise ratio. We fitted the RME of both nights jointly using the theoretical model provided in the software package `EXOFAST_RV` (*48*). The free parameters of this fit are: the mid-transit time $T_0$, a linear limb darkening coefficient $u_1$, the spin-orbit misalignment ($\lambda$), and the equatorial velocity of the star times the sine of the inclination of the stellar axis ($v \sin i$). We also allow for a linear RV drift in time. The RV data of both nights and the respective best fitting model are shown in Fig. S5. The best fitting mid-transit time differs from the literature prediction by 1.8 ± 1.0 min (1σ), which is consistent with the propagated error of the literature ephemeris (*14*). Using the mid-transit time derived from the RME fit instead of the literature value has negligible impact on our results. Our derived best fitting parameters are $T_0$ = 2457988.48748 ± 0.00070 $BJD_{TBD}$, $u_1$ = 0.9 ± 0.1, $v \sin i = 1.8 \pm 0.1$ km s$^{-1}$, $\lambda = 1.2^\circ \pm 1.9^\circ$. We discuss the potential effect of the stellar line distortion by the RME on the transmission spectrum below.

Calculation of the equivalent height of the absorbing atmosphere $\delta_{Rp}$

The excess depth measured in an atmospheric absorption line is often described as a wavelength dependent planetary radius. The increase in transit depth $\Delta d$ due to such a planet radius larger than the continuum radius by $\delta_{Rp}$ is given by

$$\Delta d = ((R_p + \delta_{Rp})^2 - R_p^2)/R_s^2 \qquad \text{(S1)}$$

As we measure $\Delta d$, we can calculate the equivalent height of the absorbing atmosphere as

$$\delta_{Rp} = (\Delta d\, R_s^2 + R_p^2)^{1/2} - R_p, \qquad \text{(S2)}$$

where $R_p$ is the planet radius and $R_s$ the stellar radius. We are using the term 'equivalent height' for $\delta_{Rp}$, as physically, the absorbing atmosphere does not need to present as a homogeneously increased planet radius but can, for example, be elongated into a tail like feature, which appears to be the case for WASP-69b.

Estimation of contamination of the transmission spectrum by the Rossiter-McLaughlin effect

The stellar center-to-limb variation (CLV) and the Rossiter-McLaughlin effect both cause a distortion in the stellar line shape during transit, which is known to imprint itself on potential transmission signals (*15*, *22, 55*).

The RME in the helium line is expected to exhibit a slightly different shape, amplitude and duration compared to the RME in other stellar lines due to the increased absorption of the extended helium atmosphere and the tail. The exact orientation of the tail and, thus, the part of the stellar disk occulted by it at a given moment is unknown. However, we can estimate the potential impact of stellar line distortion on the derived helium transmission spectrum by modeling the effect with the RME caused by a crossing object larger than the broadband planet (i.e. radius = 1.73 $R_p$). We model the stellar line distortion with the method described in (*22, 55*). We simulate the stellar lines using Spectroscopy Made Easy tool (*56*) with the MARCS model (*57*). The stellar parameters used are the effective temperature $T_{eff}$ = 4700 K, the logarithm of the gravitational acceleration log $g$ = 4.5 and the metallicity [M/H] = 0.15. The stellar model cannot simulate the He I lines as the lines do not originate in the photosphere, so we use the actually observed and normalized He I lines (obtained during out-of-transit) as an input for the model. By applying this method, we can simulate the RME effect of the He I lines but not the CLV effect. For other stellar lines in the wavelength region both RME and CLV are modeled. For the helium line RME we repeat the simulation with the inflated effective planet radius of 1.73 $R_p$, (i.e. 1 $R_p$ + $\delta_{Rp}$). In all cases we find the expected change to the transmission spectrum to be negligible. When considering the planet size as 1.00 $R_p$ the change in depth of the He I feature is ~ 0.03% and ~ 0.07% for an object with 1.73 $R_p$. For 1.73 $R_p$ the measured blue shift further changes by +0.13 km $s^{-1}$. We also consider the effect of line distortion for the other investigated planets. For HD 189733 we repeat the same modeling as for WASP-69 and find the potential change in depth to be 0.06%. For KELT-9 and HD 209458 the effect of RME and CLV can be neglected, as we measure no stellar He I lines for either of the two stars. For GJ 436, the stellar He I lines are present, but for this system the amplitude of the RME is extremely small, i.e. of the order of 1.3 m $s^{-1}$ (*58*), which renders the impact of the RME minimal.

Calculation of the atmospheric scale height for the lower atmosphere

The strength of measurable absorption features scales with a planet's atmospheric scale height, which in general terms, is the distance in which the density decreases by a factor of *e*, the base of natural logarithms. To enable the comparison of atmospheric signals from different planets they are often shown normalized over $H_{eq}$, the planet's scale height of the lower atmosphere in hydrostatic conditions (*59*). Under these conditions it is calculated as,

$H_{eq} = (k_B T_{eq}) / (\mu\, g)$ (S3)

using the planet equilibrium temperature $T_{eq}$, the mean molecular mass $\mu$, the gravitational acceleration $g$ and $k_B$, the Boltzmann constant. We calculate $H_{eq}$ for all planets of our sample assuming $\mu = 2.3 \times m_H$, with $m_H$ the mass of a hydrogen atom, which corresponds to a solar composition hydrogen-helium mixture. The resulting values are given in Table S2.

Calculation of X-ray and extreme ultraviolet flux

The X-ray and EUV (< 50.4 nm) flux used in Table S4 and Figure 4 has been calculated using a coronal model for the sources (*60*). In brief, observations of the targets in X-rays and EUV provide spectral information that is used to calculate the amount of emitting material (emission measure) in different ranges of temperature $T$ in the transition region and corona of the star (log $T$ (K) between 4 and 7.5). This allows us to construct a coronal model, which is used to determine a synthetic spectrum in the whole X-ray and EUV spectral range (0.5-92 nm). In the cases with no ultraviolet observations, the cooler part of the coronal model is calculated by extrapolating from the hotter component (*60*).

This method has been tested and provides the EUV fluxes which cannot be observed due to interstellar absorption (*60*). The atomic database AtomDB ver. 3.0.9 is used in the spectral fitting and synthetic models (*61*).

The analysis made use of the XMM-Newton observations with the following IDs:

WASP-69: 0783560201; HD 189733: 0506070201, 0600970201, 0672390201, 0690890201, 0692290201, 0692290301, 0744980201, 0744980301, 0744980401, 0744980501, 0744980601, 0744980801, 0744980901, 0744981001, 0744981101, 0744981301, 0744981201, 0744981401, 0744980701, 0748391401, 0744981501, 0744981601, 0744981701, 0748391501; HD 209458: 0404450101; GJ 436: 0556560101, 0764100501; HAT-P-11: 0764100701.

For WASP-107 we performed new observations with XMM-Newton under Director Discretionary Time (obs ID: 0830191901, P.I. Sanz-Forcada) on 2018/06/22. The X-rays and EUV fluxes received by WASP-107b are supportive of our findings (see Table S4). The X-ray data were reduced following the standard XMM-Newton tasks in the Scientific Analysis System (SAS) software. The spectra of the three EPIC (European Photon Imaging Camera) detectors were obtained from a circle centered at the target's position, and the background from a nearby source-free region. The spectral contribution of the source was then fitted simultaneously to the background spectral contribution, to identify the net emission from the source following (*60*).

Transmission spectra of GJ 436b, HD 209458b, and KELT-9b flux

We also analyzed observations of the three planets, GJ 436b, HD 209458b, and KELT-9b, in the same way as the CARMENES observations of WASP-69b. The observation for KELT-9b was performed on 2017 August 6. The observation lasted for ~ 6 hours including ~ 2 hours out-of-transit time. The observation for GJ 436b was conducted on 2018 February 14, with baseline observations starting 5.78 hours before the transit ingress and covering 1.16 hours after transit. The observations for HD 209458b were obtained on 2016 September 16 and 2016 November 8, lasting ~6.3 hours and 5 hours, including ~3.3 hours and 2 hours of out-of -transit baseline, respectively. None of the data sets show clear signs for helium absorption (Fig. S10). Based on the standard deviation of the spectra we derive 90% confidence upper limits for KELT-9b of 0.33% and 0.41% for GJ 436b. The HD 209458 datasets exhibit relatively low white noise but red noise hampers the search for a possible He I signal. We determine a 90% upper limit of 0.84% for this planet. Details on the He I detection in HD 189733b and the corresponding transmission spectrum are presented in (*24*).

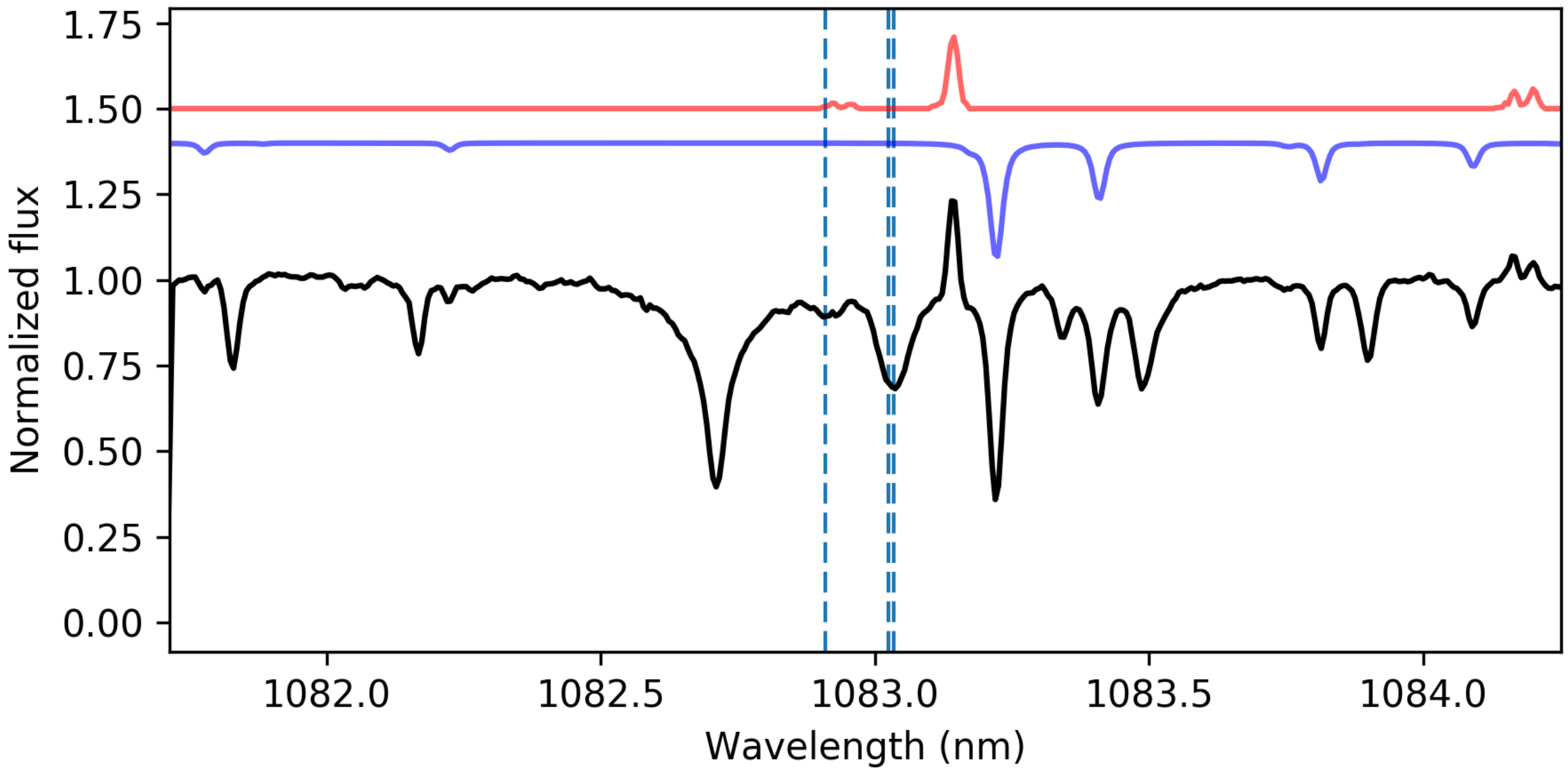

**Fig. S1. Stellar spectrum of WASP-69 showing the He I region.** In black we show an example stellar spectrum of the region around the He I triplet. The positions of the three lines comprising the triplet are indicated as dashed vertical lines. The model for telluric water vapor absorption is plotted in blue, and the model for the emission lines is shown in red.

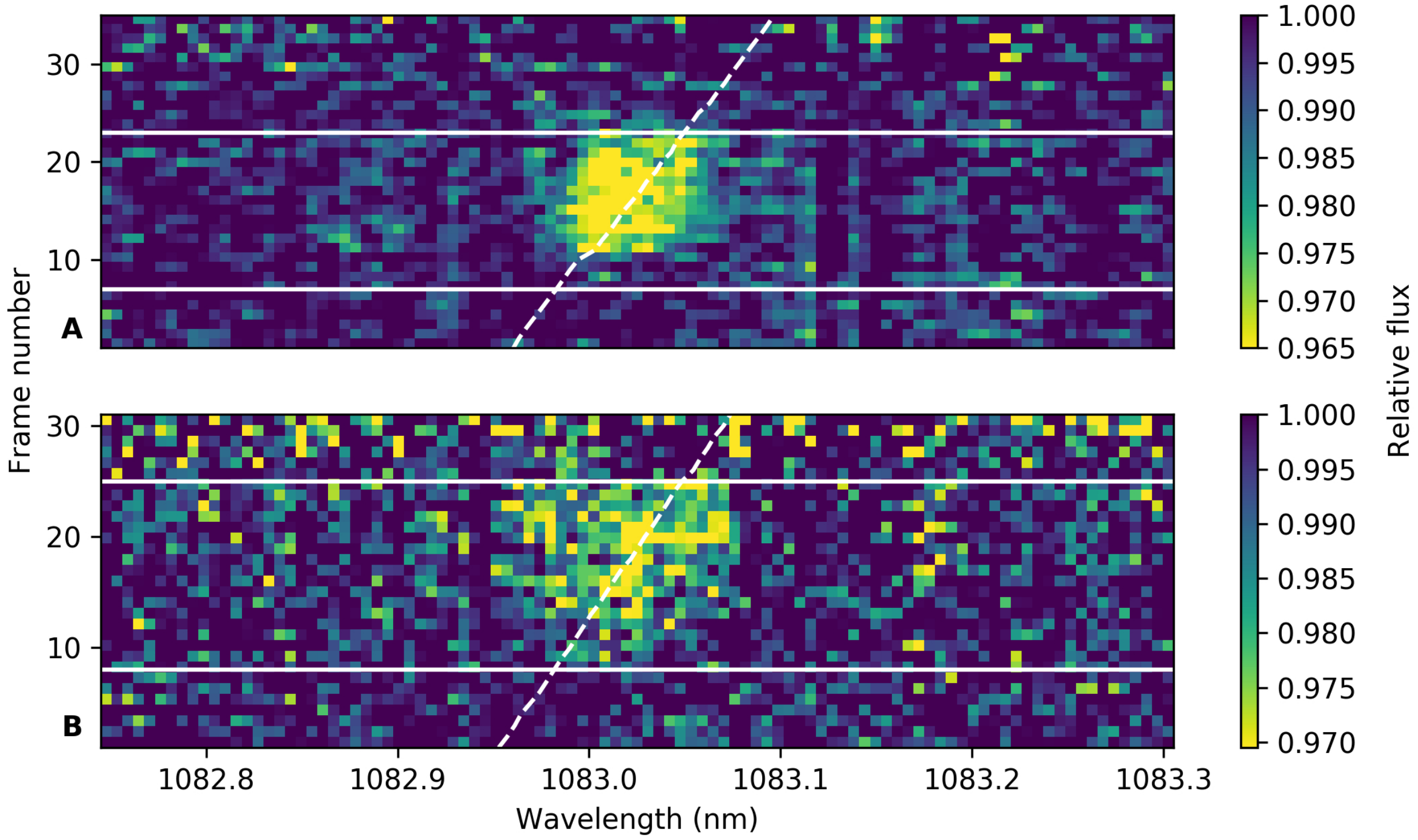

**Fig. S2. The planetary helium absorption in every spectrum of night 1 (panel A) and night 2 (panel B).** Lighter color indicates deeper absorption. We mark the first and fourth contact of the broadband transit (white lines) and the predicted planetary velocity shift due its orbital motion (white dashed line).

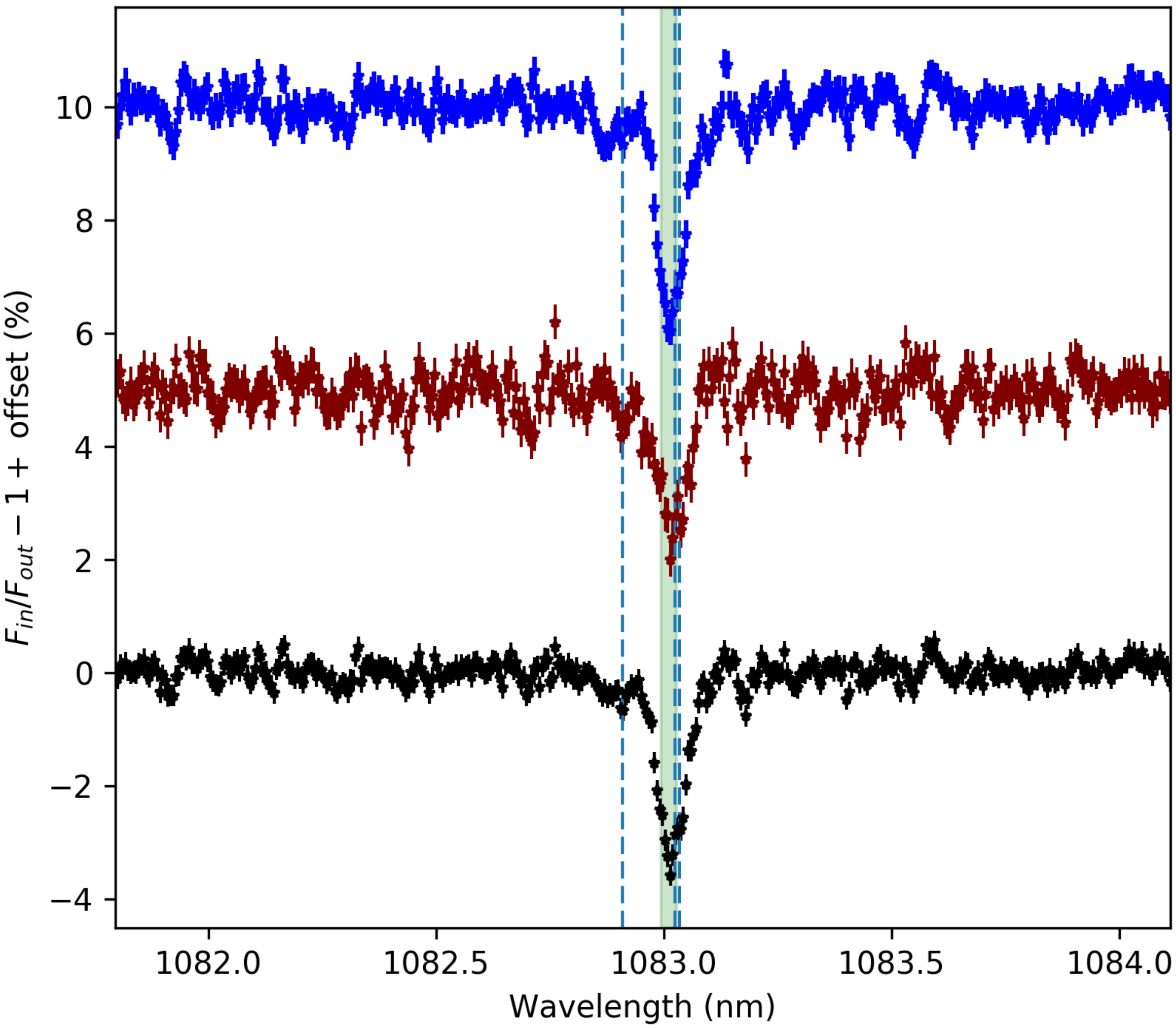

**Fig. S3. Transmission spectra for each individual night, and for the combined spectrum.** The uppermost transmission (blue) was obtained in night 1, the middle one (maroon) in night 2 and the lowest spectrum (black) represents the weighted mean of both nights (as shown in Fig. 2). The He I triplet exhibits a blue shift from its rest position (vertical dashed lines) in both nights. The 0.04 nm wide passband used to build the light curves in Fig. 3 is indicated as a green shaded area.

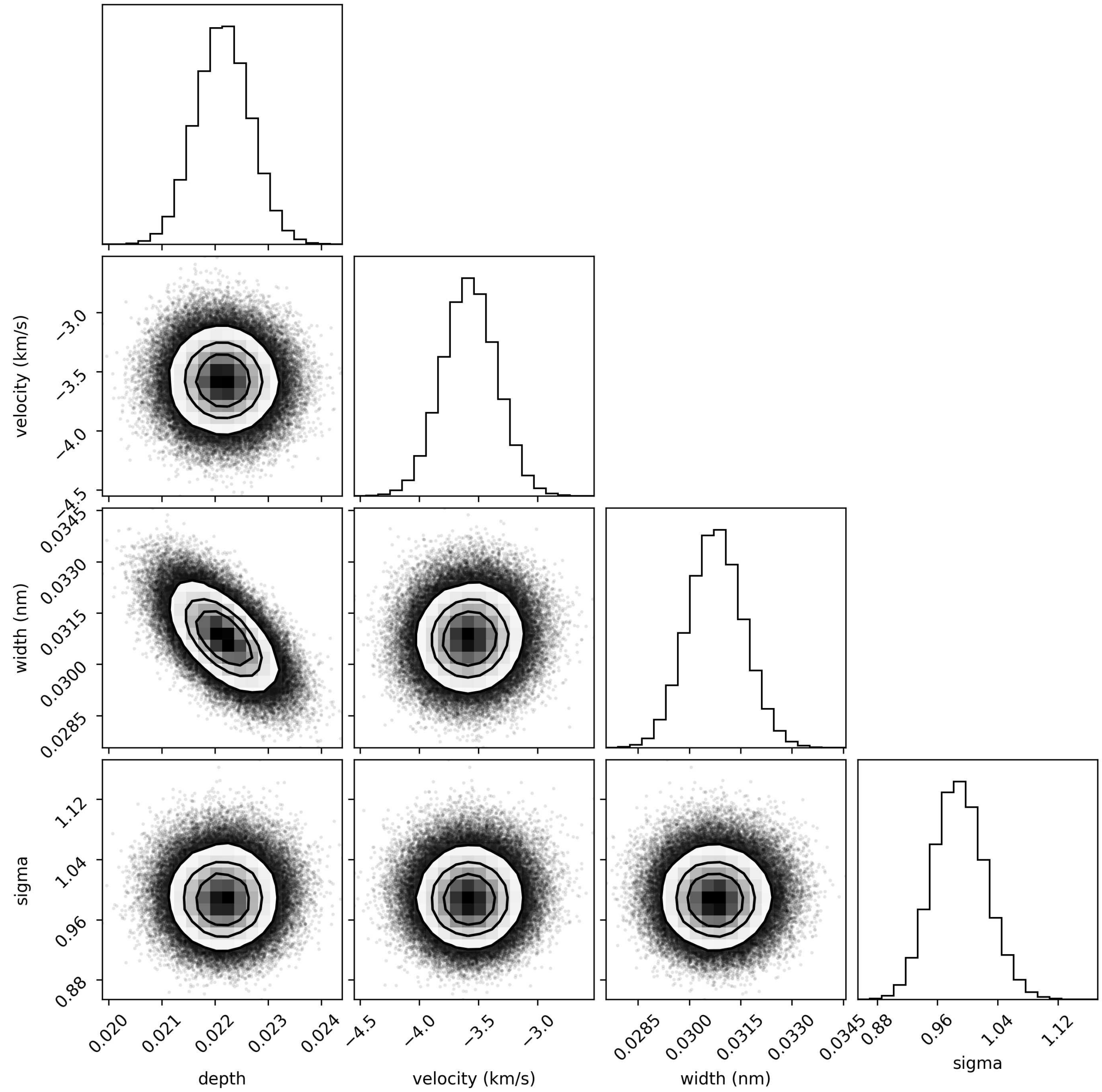

**Fig. S4. Marginal and joint posterior distributions for the retrieved parameters for the transmission spectrum.** The contours indicate the areas encompassing 68.27% (1σ), 95.45% (2σ) and 99.73% (3σ) of the posterior probability. The low density region beyond 3σ is indicated by plotting the individual points of the distribution in black. This plot was made using the python module corner.py (*62*).

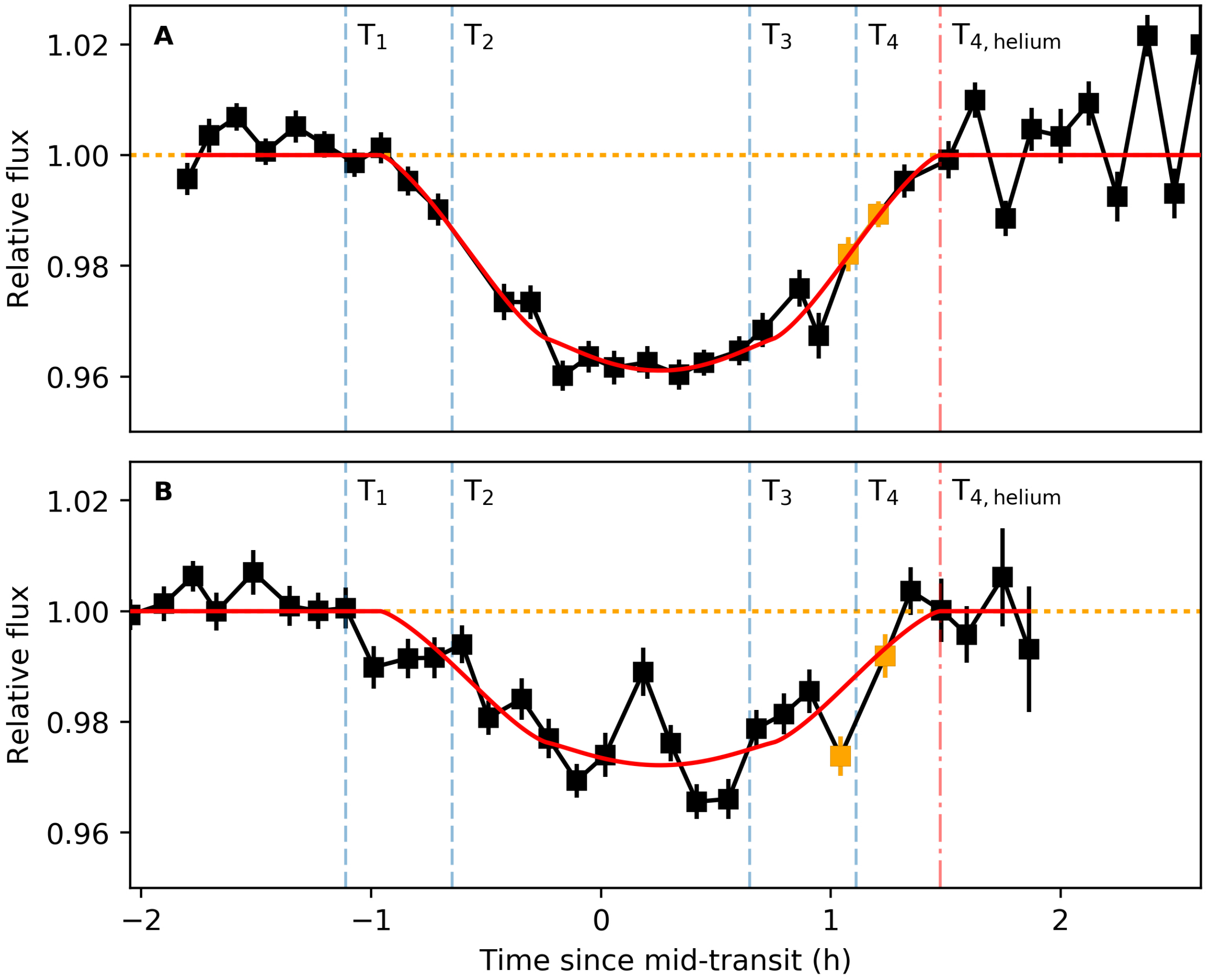

**Fig. S5. Fitted light curve models.** The data of both nights are shown together with the fitted light curves (red line) used to determine the time of $T_{4,helium}$. The two spectra of each night used to build the $T_4$-spectrum i.e. the 'tail-spectrum' shown in Fig. S9 are marked in orange.

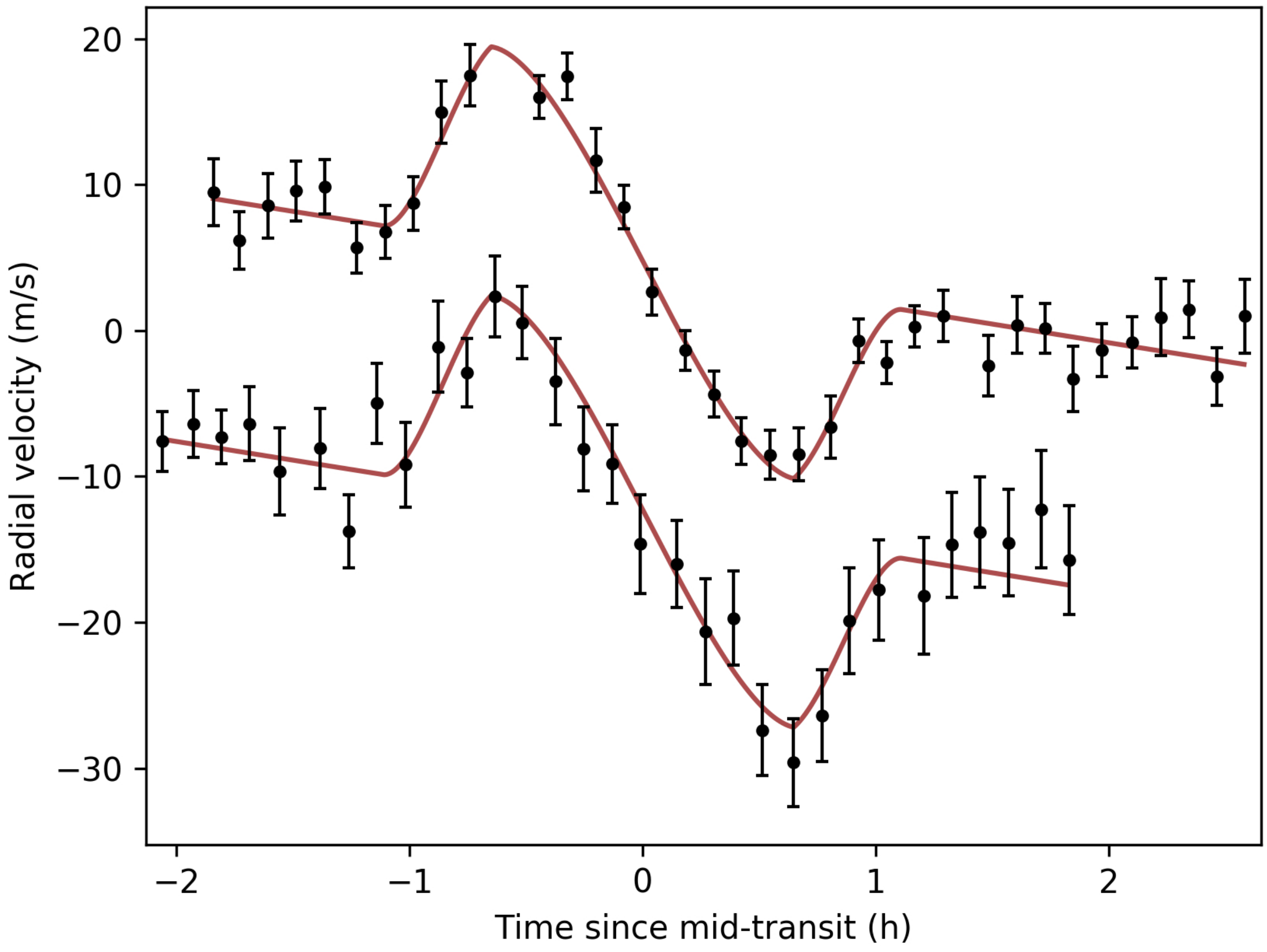

**Fig. S6. Radial velocity curves of WASP-69 showing the broadband Rossiter-McLaughlin effect.** We show the data obtained in the visible-light channel of CARMENES after correction of a linear drift and shifted by a constant offset for clarity. The upper curve represents night 1 and the lower curve night 2. The best fitting model is overplotted in red.

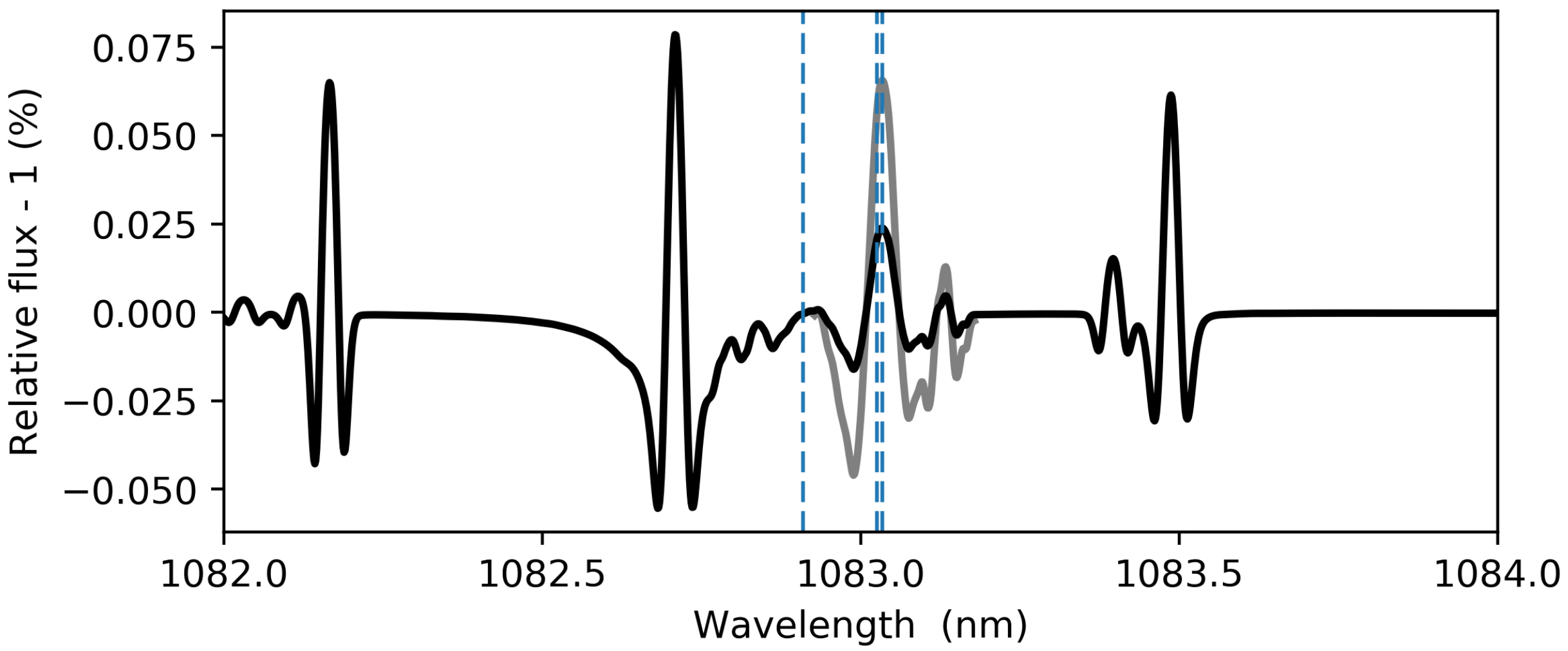

**Fig. S7. Estimated signal of the RME in the transmission spectrum.** The effect is modeled for an object with a radius of 1.00 $R_p$ (black curve) and also for an object with a radius of 1.73 $R_p$ in the He I lines (grey curve) using the Spectroscopy Made Easy tool. The tool cannot simulate the He I lines as they originate in the stellar chromosphere, thus, only the RME is considered for the line. All other stellar lines include both RME and CLV modeling.

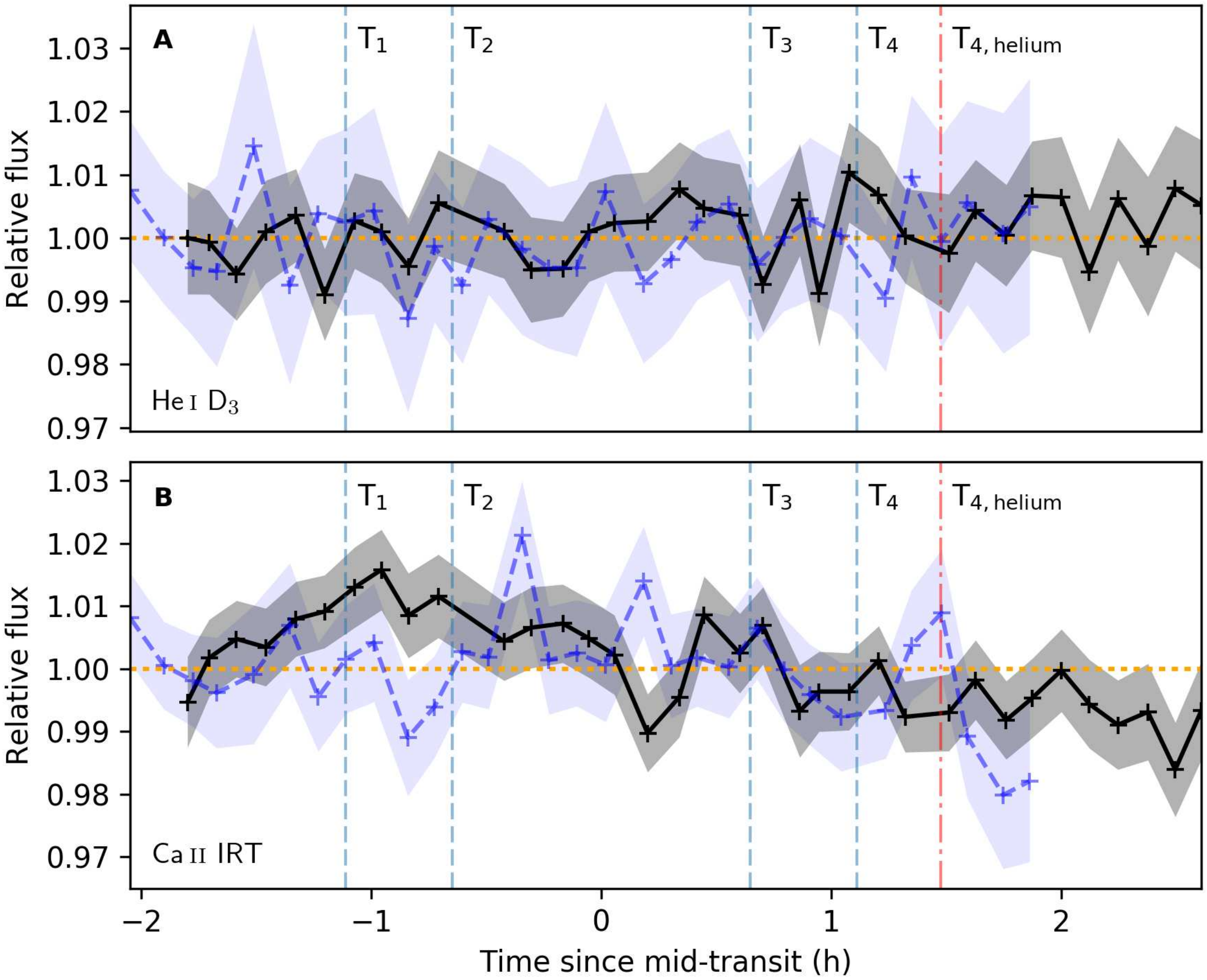

**Fig. S8. Light curves centered on the core of the He I $D_3$ line (A) and the cores of the three the Ca II IRT lines (B).** The light curves of the indicator lines were built using 0.04 nm wide bins centered on the respective line cores in the stellar rest frame. Neither the He I $D_3$ line at 587.6 nm nor the Ca II IRT lines at 849.8 nm, 854.2 nm and 866.2 nm show indications of occulted active regions. There is no correlation of the curves with the time of the planet transit, which would be indicative of unocculted active regions.

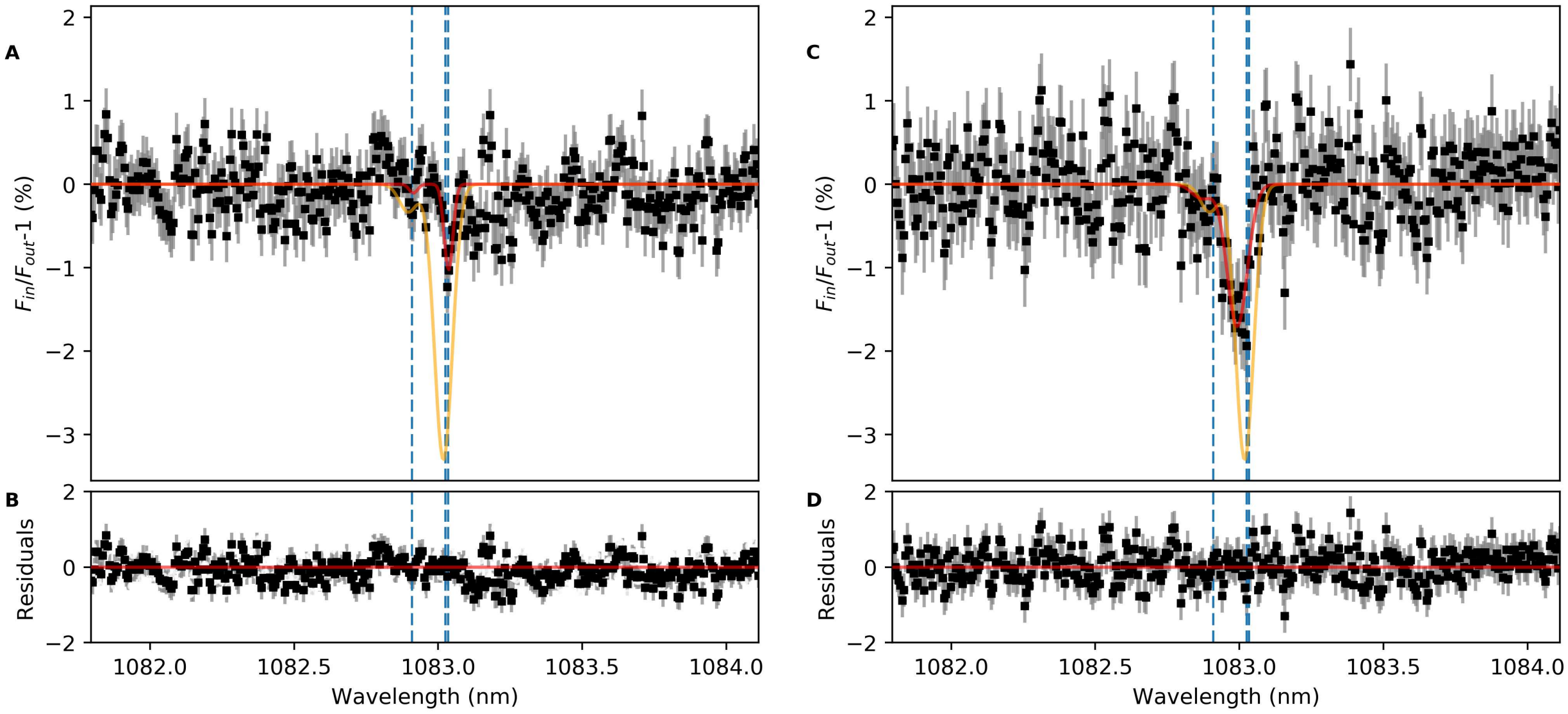

**Fig. S9. Weighted mean spectrum obtained between first and second contact (A, B) and close to the time of fourth contact (C, D).** The spectrum shown in A was constructed using the spectra of each night obtained between $T_1$ and $T_2$, i.e. during the ingress of the broadband transit. This ‘ingress’ spectrum exhibits a lower amplitude than the $T_2$-$T_3$ spectrum and is slightly red shifted instead of blue shifted. The best fitting model for the $T_1$-$T_2$ spectrum is plotted in red. The spectrum shown in C was constructed using the two spectra of each night that were obtained closest to $T_4$, when the broadband transit has ended and only the helium tail occults the stellar disc (see Fig. S5). The tail spectrum exhibits a lower amplitude and larger blue shift than the $T_2$-$T_3$ spectrum. The best fit model for the $T_4$ spectrum is plotted in red. For comparison the best fitting model for the $T_2$-$T_3$ spectrum is plotted in yellow in A and C.

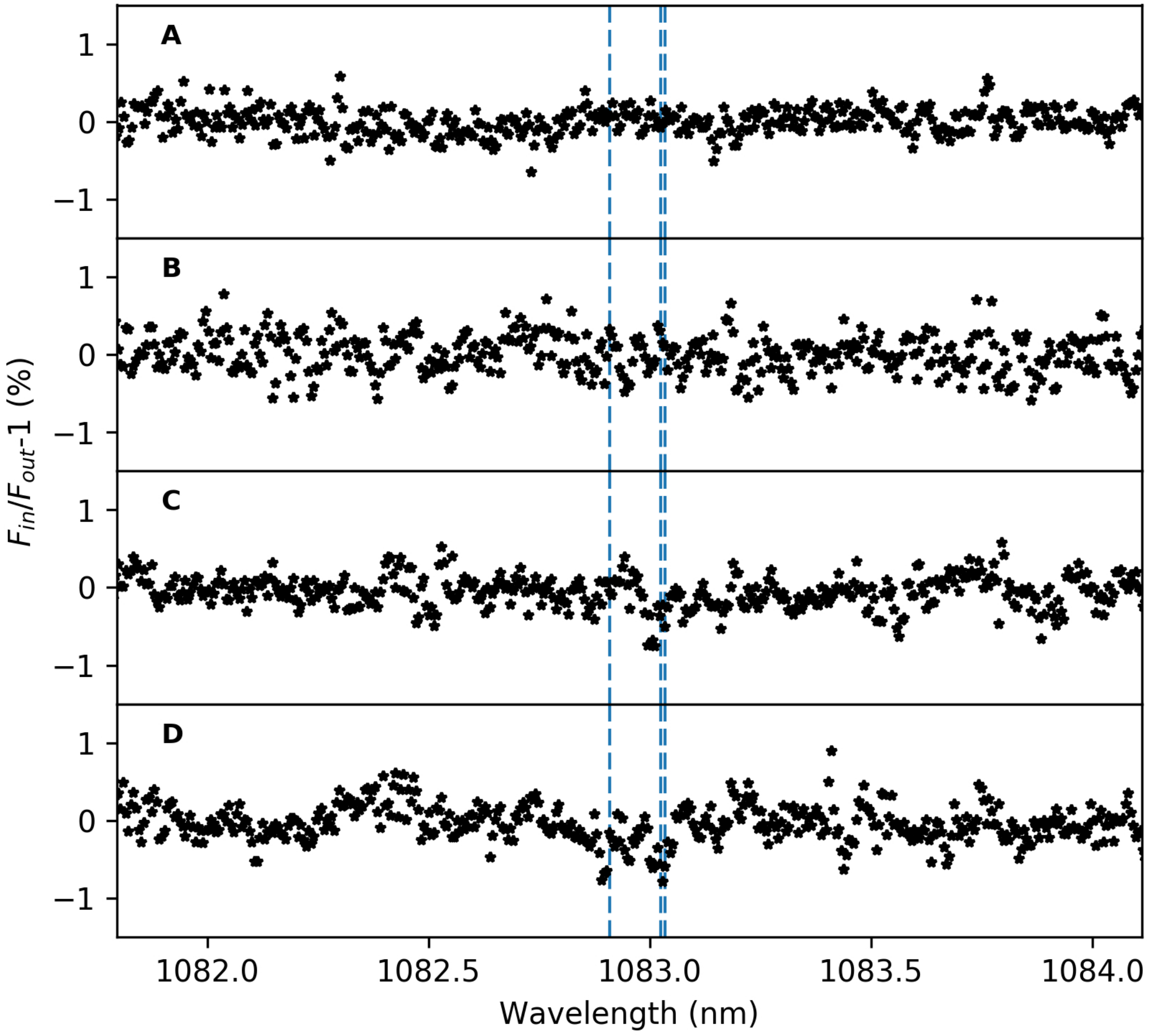

**Fig. S10. Transmission spectra obtained for KELT-9b (A), GJ 436b (B) and for two different observing opportunities for HD 209458b (C, D).** Vertical dashed lines indicate the theoretical wavelengths of the He I triplet lines. Equivalent data for HD 189733b are presented in (*24*).

**Table S1. Observing log of the CARMENES transit observations of WASP-69b.** We list the starting and end time of the observations, the minimal and maximal airmass during the observations, the average seeing during the night and the average signal-to-noise ratio (S/N) achieved for an individual exposure. The last three columns give the number of spectra obtained between the second and third and between the first and fourth contact of the planet transit, as well as the number of spectra obtained out-of-transit.

| Date | Start (UT) | End (UT) | Min. airmass | Max. airmass | Av. seeing in V band (arcsec) | Average S/N | # spectra ($T_2$-$T_3$) | # spectra ($T_1$-$T_4$) | # spectra out of transit |
|---|---|---|---|---|---|---|---|---|---|
| 22 Aug 2017 | 21:40 | 02:06 | 1.35 | 1.97 | 0.63 | 77 | 9 | 17 | 18 |
| 22 Sep 2017 | 20:20 | 00:04 | 1.35 | 1.97 | 0.69 | 67 | 10 | 18 | 13 |

**Table S2. Physical parameters and calculated atmospheric scale heights for the lower atmosphere.** $T_{eq}$ is the planet equilibrium temperature and $g_p$ is the planet's surface gravity. Planet and stellar radius are given in units of Jupiter radius ($R_{Jup}$= 7.1492×$10^7$ m) and Solar radius ($R_{Sun}$= 6.95508×$10^8$ m). Upper limits correspond to 90% confidence.

| System | $T_{eq}$ (K) | $g_p$ (m $s^{-2}$) | $R_p$ ($R_{Jup}$) | $R_s$ ($R_{Sun}$) | $H_{eq}$ (km) | $\delta_{Rp} / H_{eq}$ | $\delta_{Rp}$ /($R_p$) | Ref. |
|---|---|---|---|---|---|---|---|---|
| WASP-69 | 963 | 5.32 | 1.057 | 0.813 | 650 | 85.5 ± 3.6 | 0.73 ± 0.03 | (*14*) |
| HD 189733 | 1201 | 19.50 | 1.216 | 0.805 | 221 | 77.2 ± 6.1 | 0.20 ± 0.02 | (*63-65*) |
| HD 209458 | 1449 | 7.59 | 1.451 | 1.203 | 658 | ≤ 36.9 | ≤ 0.24 | (*63, 66*) |
| GJ 436 | 675 | 13.18 | 0.369 | 0.455 | 184 | ≤ 37.5 | ≤ 0.26 | (*67*) |
| KELT-9 | 4050 | 19.95 | 1.891 | 2.362 | 729 | ≤ 41 | ≤ 0.22 | (*68*) |

**Table S3. Equatorial coordinates for all systems.** Given are right ascension (RA) and declination (Dec.) from (*69*). Also listed are the references for planet discovery.

| System | RA (J2000) | Dec. (J2000) | Discovery |
|---|---|---|---|
| WASP-69 | $21^h00^m06^s.197$ | -05°05'40".04 | (*14*) |
| HD 189733 | $20^h00^m43^s.713$ | +22°42'39".07 | (*70*) |
| HAT-P-11 | $19^h50^m50^s.247$ | +48°04'51".10 | (*71*) |
| HD 209458 | $22^h03^m10^s.773$ | +18°53'03".55 | (*72*, *73*) |
| GJ 436 | $11^h42^m11^s.093$ | +26°42'23".66 | (*27*) |
| WASP-107 | $12^h33^m32^s.844$ | -10°08'46".22 | (*74*) |
| KELT-9 | $20^h31^m26^s.353$ | +39°56'19".77 | (*68*) |

**Table S4. Host star activity and calculated X-ray and EUV flux and for the systems observed for planetary He I absorption.** The EUV values take into account photons with wavelength 10.0 - 50.4 nm (instead of the usual range 10 – 92 nm), i.e. only those with energies high enough to ionize helium. The first three columns give the distance (d) of each star to Earth, the measured X-ray (0.5 – 10 nm) luminosity ($L_X$) and the EUV luminosity ($L_{EUV}$). These calculations are based on coronal models for each of the host stars, based in turn on X-ray and UV observations of the targets following (*60*), with the exception of KELT-9 for which we have no X-ray/EUV observations. The A0V-type host star KELT-9 has no expected corona and, thus, its X-ray/EUV contribution can be approximated as blackbody emission from the photosphere. The $F_X$ and $F_{EUV}$ columns give the X-ray and EUV incoming flux at the orbital distance of the planet from the star, respectively. The uncertainties given for the EUV values are propagated from the uncertainties in the coronal modeling and the ones given for X-ray values are derived from the signal-to-noise ratio of the XMM observations. The last column gives the value for the host star activity indicator log(R'$_{HK}$). The values for log(R'$_{HK}$) are taken from (*14*) for WASP-69b and (*29*) for the remaining stars. For KELT-9 and WASP-107 no measurement of this index exists.

| Host star | d (pc) | $L_X$ (× $10^{20}$ W) | $L_{EUV}$ (× $10^{20}$ W) | $F_X$ (W/m$^2$) | $F_{EUV}$ (W/m$^2$) | log(R'$_{HK}$) |
|---|---|---|---|---|---|---|
| WASP-69 | 50.3 | 14.50 ± 0.47 | 9.51 ± 3.24 | 2.520 ± 0.082 | 1.650 ± 0.560 | -4.54 |
| HD 189733 | 19.3 | 20.20 ± 0.05 | 26.30 ± 0.06 | 7.280 ± 0.018 | 9.470 ± 0.021 | -4.501 |
| HAT-P-11 | 37.5 | 3.53 ± 0.24 | 13.10 ± 0.97 | 0.446 ± 0.033 | 1.663 ± 0.120 | -4.567 |
| HD 209458 | 47.0 | 0.36 ± 0.14 | 1.08 ± 0.64 | 0.057 ± 0.022 | 0.170 ± 0.100 | -4.970 |
| GJ 436 | 10.2 | 0.15 ± 0.01 | 0.31 ± 0.01 | 0.066 ± 0.003 | 0.131 ± 0.006 | -5.298 |
| WASP-107 | 64.9 | 4.83 ± 0.21 | 17.84 ± 9.00 | 0.567 ± 0.025 | 2.097 ± 1.050 | n.a. |
| KELT-9 | 206 | n.a. | 0.52 | n.a. | 0.15 | n.a. |